\documentstyle[psfig]{l-aa}

\def\mincir{\raise -2.truept\hbox{\rlap{\hbox{$\sim$}}\raise5.truept
\hbox{$<$}\ }}
\def\lsimeq
{\hbox{\raise0.5ex\hbox{$<\lower1.06ex\hbox{$\kern-1.07em{\sim}$}$}}}
\def  \magcir{\raise -2.truept\hbox{\rlap{\hbox{$\sim$}}\raise5.truept
\hbox{$>$}\ }}

\begin{document}

\thesaurus{ 03(11.11.1;       
	    11.19.2;          
	    11.19.6)}         


\title{{\it BeppoSAX} detection of the Fe K line in the nearby starburst
galaxy NGC 253
} 


\author{
	M.~Persic\inst{1},
	S.~Mariani\inst{2},
	M.~Cappi\inst{3},
	L.~Bassani\inst{3},
	L.~Danese\inst{4},
	A.J.~Dean\inst{5},
	G.~Di~Cocco\inst{3},
	A.~Franceschini\inst{6},
	L.K.~Hunt\inst{7},
	F.~Matteucci\inst{8},
	E.~Palazzi\inst{3},
	G.G.C.~Palumbo\inst{2,3},
	Y.~Rephaeli\inst{9},
	P.~Salucci\inst{4}, and 
	A.~Spizzichino\inst{3}}

\offprints{Massimo~Persic; e-mail:{\tt persic@oat.ts.astro.it}}

\institute{
Trieste Astronomical Observatory, via G.B.Tiepolo 11, 
	34131 Trieste, Italy 					\and 
Astronomy Dept., University of Bologna, via Zamboni 33, 
	40126 Bologna, Italy 					\and  
ITeSRE/CNR, via Gobetti 101, 40129 Bologna, Italy 		\and
SISSA/ISAS, via Beirut 2-4, 34013 Trieste, Italy  		\and 
Physics Dept., Southampton University, Southampton SO9 5NH, UK  \and
Astronomy Dept., University of Padova, vicolo dell'Osservatorio 5, 
	35122 Padova, Italy  					\and
CAISMI/CNR, Largo E.Fermi 5, 50125 Firenze, Italy 		\and 
Astronomy Dept., University of Trieste, via Besenghi, 
	34131 Trieste, Italy 					\and
School of Physics and Astronomy, Tel Aviv University, Tel Aviv 69978,
	Israel}

\date{Received..................; accepted...................}

\maketitle
\markboth{Persic et al.: X-Ray Spectrum of NGC~253}{}

\begin{abstract}
We present {\it BeppoSAX} results on the nearby starburst galaxy NGC 253.
Although extended, a large fraction of the X-ray emission comes from the
nuclear region. Preliminary analysis of the LECS/MECS/PDS 
$\sim$ 0.2-60 keV data from the central 
4$^{\prime}$ region indicates that 
the continuum is well fitted by two thermal models: a
``soft'' component with $kT \sim 0.9$ keV, and a ``hard'' 
component with $kT \sim 6$ keV absorbed by a column density of $\sim$1.2
$\times$ 10$^{22}$ cm$^{-2}$. For the first time in this object, the Fe K
line at 6.7 keV is detected, with an equivalent width of $\sim$300 eV.
This detection, together with the shape of the 2--60 keV continuum, implies
that most of the hard X-ray emission is thermal in origin, and constrains
the iron abundances of this component to be $\sim$0.25 of solar. Other
lines clearly detected are Si, S and Fe L/Ne, in agreement with previous
{\it ASCA} results. We discuss our results in the context of the
starburst-driven galactic superwind model. 
\end{abstract}


\section{Introduction}

In the local universe, starburst galaxies (SBGs) are galaxies with current
(for typically $10^{8}$ yr: Rieke et al. 1980) star formation rates
significantly higher than the galactic average. In the early evolution of
galaxies ($z \magcir 1$), star formation was certainly much higher than at
present (see e.g. Madau et al. 1996; Hogg et al. 1998): a starburst phase
was, therefore, normal in the early evolution of galaxies. Part of the
interest in local SBGs stems then from the possibility that they resemble
normal galaxies when they were at $z \magcir 1$.  High-energy phenomena
highlight the SB activity (Bookbinder et al. 1980): because of obscuration
of the optical emission (by star-heated IS dust), X-rays allow a
penetrating view of SBGs. 

Aiming at studying the X-ray properties of local SBGs, we have undertaken
a program to study two such sources, NGC 253 and M82, with {\it BeppoSAX}:
in this Letter the broad-band spectrum of the nuclear region of NGC 253 is
reported, focusing on the discovery of the Fe K line at 6.7 keV and on the
features of the continuum in the 2--10 keV band arising from the nuclear
region. 

Assumed to be a prototypical SBG, NGC 253 has been extensively observed in
X-rays.  At low energies, analysis of Einstein IPC data led Fabbiano
(1988) to conclude that the nuclear emission was softer and more absorbed
than that from the entire galaxy. A subsequent observation by {\it Ginga}
in the 2--10 keV band (Ohashi et al. 1990) indicated that the spectrum was
best fitted by a 6-7 keV thermal bremsstrahlung and no significant
absorption; no iron line emission was detected ($EW <$400 eV). A more
recent {\it ASCA} measurement (Ptak et al. 1997) required a two-component
model to explain the complexity of the NGC 253 X-ray spectrum [i.e., a
soft thermal component with $kT \sim$ 0.8 keV, plus a hard absorbed
($N_{\rm H} \sim $ 10$^{22}$ cm$^{-2}$) component, described either as
thermal ($kT \sim$ 7 keV) or power-law ($\Gamma \sim 2$)]: again, no iron
line emission was detected ($EW <$180 eV). At higher energies, a 4.4
$\sigma$ detection by {\it OSSE} has been claimed by Bhattacharya et al.
1994 with a 50--165 keV flux of 3 $\times$ 10$^{-11}$
ergs~cm$^{-2}$~s$^{-1}$, higher than that extrapolated from the {\it
Ginga}/{\it ASCA} observation.

\section{The {\it BeppoSAX} observation}

NGC 253 was observed on Nov.29--Dec.2, 1996 (see Table 1).  The source is
clearly extended (and elongated) up to $\sim$ 10$^{\prime}$ in the {\it
BeppoSAX} image in both the 0.1--2 keV and 2--10 keV band.  Analysis of
the resolved emission is deferred to a separate paper (but see preliminary
results in Cappi et al. 1998):  here we only present the analysis of the
unresolved nuclear emission. In the central 4$^{\prime}$ region (typical
extraction region for point sources used for the LECS and MECS instruments), 
we obtained a LECS count rate of $3.93 \times 10^{-2}$~cts~s$^{-1}$ and a 
MECS count rate of $9.23 \times 10^{-2}$ cts~s$^{-1}$. Background spectra, 
extracted from the standard blank-sky files provided by the $BeppoSAX$ 
Science Data Center (SDC), contributed less than about 5\% and 15\% at 2 and 
6 keV, respectively: similar results were also obtained using background 
spectra extracted directly from the instruments' fields of view. LECS data 
above 4 keV were excluded because of remaining calibration uncertainties. 
LECS and MECS data were reduced using the SAXDAS v.1.3.0 and XANADU 
packages, while the XAS v.2.0 package was used for the PDS data. LECS and 
MECS spectra were rebinned in order to obtain at least 20 counts per energy 
channel, to allow use of the $\chi^2$ statistics. 


\begin{table}
\begin{center}   
\begin{tabular}{ccc}
\multicolumn{3}{c}{Table 1: Exposure times.} \\
\hline
Instrument &  En. range & Obs. time \\
$ $ & $keV$ & $sec$ \\
\hline
LECS & 0.1-4 & 54689 \\
MECS & 1.3-10 & 113403 \\
PDS  & 13-60  & 51557   \\
\hline
\end{tabular}
\end{center}
\end{table}

No short or long term variability is detected from the present data in any
energy band: a fact that can be simply interpreted as evidence for the
large size of the emitting region.  The 0.5--2 keV and 2.0--10 keV fluxes
observed by {\it BeppoSAX} are $\sim$2.53$\, \times
10^{-12}$~erg~s$^{-1}$~cm$ ^{-2}$ and 4.9 $\times 10^{-12}$~erg~s$^{-1}
$~cm$^{-2}$. These are roughly consistent with the observed ROSAT PSPC and
{\it ASCA} fluxes (Serlemitsos et al. 1996; Ptak et al. 1997) if one takes
into account the different sizes of the source regions. The global,
absorption-corrected, 0.5--2 keV plus 2.0--10 keV luminosities are 7.72
$\times$ 10$^{39}$ ergs~s$^{-1}$ and 1.57 $\times$ 10$^{40}$ ergs~s$^{-1}$
for the soft and the hard component (see Table 3), respectively, for an
assumed distance of 3 Mpc. 

The {\it BeppoSAX} PDS count-rate between 13--60 keV is 0.07$\pm$0.02
cts~s$^{-1}$ which corresponds to a $\sim$2.5 $\sigma$ level detection if
systematics errors ($\sim$ 10\%) are taken into account: thus the
3$\sigma$ upper-limit to the PDS 50--165 keV flux is 7.1 $\times$
10$^{-12}$ ergs cm$^{-2}$ s$^{-1}$, much lower than the flux claimed to
have been detected by OSSE. 

In the following, $N_{\rm Hgal}$=$1.28 \times 10^{20}$ cm$^{-2}$
(Dickey \& Lockman 1990) is assumed.

\begin{table}
\begin{center}
\begin{tabular}{ccc}
\multicolumn{3}{c}{Table 2: Bremsstrahlung plus emission lines model.} \\
\hline
 $~~~~~$ Element ID & Obs. Energy & EW \\ 
 $~~~~~$ for K$_{\alpha}$ line & keV & eV \\ 
\hline 
&& \\
 $~~~~~$ Fe L, Ne IX/X &  $0.95_{-0.04}^{+0.03}$ & $109_{-42}^{+44}$ \\ 
&& \\
 $~~~~~$ Si XIV/XV  & $1.91_{-0.04}^{+0.04}$ & $67_{-23}^{+23}$ \\
&& \\
 $~~~~~$ S XV &  $2.42_{-0.06}^{+0.04}$ & $79_{-34}^{+34}$ \\
&& \\
 $~~~~~$ Fe XXV &  $6.69_{-0.07}^{+0.07}$ & $329_{-109}^{+89}$ \\
&& \\
\hline
\end{tabular}
\end{center}
Notes: The value of the relative normalizations $A_{LECS}/A_{MECS}$ is 
$0.64_{-0.03}^{+0.04}$. Intervals are at 90\% confidence for one interesting 
parameter.
\par\noindent
$ ^{*}$ Line energy fixed to the {\it ASCA} value (Ptak et al. 1997).
\end{table}

\begin{figure}
\psfig{file=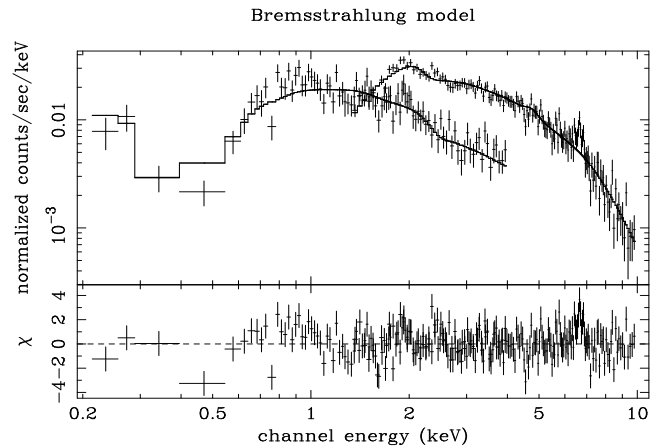,width=9.5cm,height=6.5cm,angle=-90}
\caption{Thermal bremsstrahlung fit to the MECS and LECS data; 
the lines stand out clearly on the continuum.}
\label{fig:largenenough}
\end{figure}

\begin{figure}
\psfig{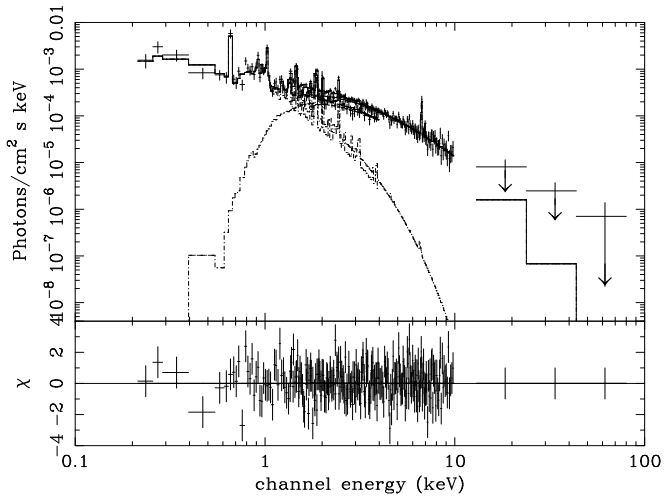}
\caption{Variable-abundances Mekal model fit to the MECS, LECS and PDS data.}
\label{fig:toosmall}
\end{figure}

\section{Spectral analysis}

We first used a thermal bremsstrahlung continuum plus emission lines model
to parameterize the line energies and intensities. The LECS and MECS
spectra were fitted simultaneously, with free relative normalization. The
best-fit value for the central temperature is $kT=7.40_{-0.71}^{+0.18}$
(with $\chi^2$/dof=254/232). A few emission lines are evident in Fig.1:
their resulting intensities and energies are listed in Table 2. In
agreement with the {\it ASCA} results, most lines are consistent with
K$_{\alpha}$ emission from H-like and He-like ions. 

At low energies ($E \mincir$ 3 keV) the strong emission lines, notably the
residual broad excess at $E<1$~keV (a clear signature of the Fe L/Ne
complex), require: {\it (a)} a soft thermal component with $kT<$1~keV; and
{\it (b)} that the hard component be absorbed in order not to overproduce
the continuum for $E\mincir$0.7 keV (see also Ptak et al. 1997). 

At higher energies, an Fe K line at $\sim$ 6.7 keV is unambiguously
detected for the first time in NGC 253. It is narrow ($\sigma < 0.13$
keV), with $EW=330$ eV and consistent with emission from
Fe XXV. [The measured equivalent width is roughly consistent with the
upper limits placed by previous studies (Ohashi et al. 1990; Ptak et al.
1997) if one considers that those authors used a flatter power-law 
continuum.] Moreover, compared with {\it ASCA}, the
larger MECS effective area at E$>$7 keV allows us to firmly establish that
the 2--10 keV continuum is dominated by thermal emission, since a hard
power law alone (still allowed by the {\it ASCA} data: Ptak et al. 1997)
yields a significantly worse fit ($\Delta$$\chi^{2}$=20) with the present
data. 

A two-temperature plasma is also required by the following consideration.
An atomic line energy is a function of the electron temperature $kT$ and
the degree of ionization (measured by $n_et$, where $n_e$ is the electron
density and $t$ the elapsed time). The centroids of the identified K-lines
(Fe, Mg, Si, S: see Table 2)), imply allowable $kT$-$n_et$ regions (see
Fig.2b of Kaneda et al. 1997) that are not compatible with a
single-temperature plasma. Thus, at least two plasma components (with
different ionization states) are required in order to reproduce the
observed SAX spectrum of NGC 253. 

Therefore our best-fit model consists of a two-component thermal model. 
Best-fit parameters are reported in Table 3 for a two-temperature
Raymond-Smith model and a two-temperature Mewe-Kaastra-Liedah plasma model
(``mekal'' in XSPEC). In order to reduce the number of free parameters,
the abundances of He, C, and N were fixed at the solar value. The heavier
elements were divided into two groups: Fe and Ni (most likely associated
with SNe I products), and the $\alpha$-elements O, Ne, Mg, Si, S, Ar and Ca
(most likely associated to SNe II products); elements in the same group
were constrained to have common abundances in solar units. 

Both components have consistent features within the statistical errors. 
The detection of the Fe K line in NGC 253 allows us to reliably determine
the Fe abundance in the line-emitting gas in the hard component: we find a
value of $\sim$ 0.25 solar, again consistent with the sub-solar values
(based on upper limits) predicted by Ohashi et al. (1990) and Ptak et al.
(1997). The iron abundance of the soft component is also constrained to a
similar value.  Instead, the abundances of the $\alpha$-elements are
consistent with a single value of $\sim$ 1.7 solar, for both spectral
components.

\section{Discussion and Conclusions}

The likely emission mechanisms and environments for X-ray emission in SBGs
have been described by Rephaeli et al. (1995). In general, they may
include thermal emission from SNe, SNRs, and hot gas in the disk and halo,
and nonthermal emission from massive X-ray binaries and Compton scattering
of relativistic electrons off the FIR and CMB fields in the disk and the
halo. Combined with the extent of the source of the 2--10 keV emission,
the detection of the spectral lines reported here is a strong evidence for
thermal hard X-ray emission from hot gas in SNRs and/or SN-driven wind,
and severely limits any contribution from a nuclear low-luminosity AGN. 

A remarkable result from the present analysis is the striking similarity
between the {\it BeppoSAX} spectrum of NGC 253 and the {\it ASCA} 0.6--10
keV spectrum of the Galactic Ridge X-Ray Emission (GRXE) in the Scutum arm
region (Kaneda et al. 1997), i.e. unresolved diffuse X-ray emission
components distributed in a thin disk along the Milky Way. 
The derived temperatures of both thermal continua and the identified lines
with their centroid energies are similar for NGC 253 and the GRXE. 

Following Kaneda et al.'s detailed analysis of the GRXE soft component, the 
departure from ionization equilibrium that can be similarly inferred in the soft 
component of NGC 253 suggests that such component may be due to SNR-related 
emission from a diffuse plasma. For an estimated SNR luminosity of $\sim 2 
\times 10^{35}$ erg s$^{-1}$ (see Kaneda et al. 1997), it can be estimated 
that the soft component's 0.5-10 keV luminosity is powered by $\sim 4 \times 
10^{4}$ SNRs. The lack of appreciable photoelectric absorption, implying a 
low HI column density where the soft component originates, also supports 
this galactic superwind picture. The derived chemical abundances of the soft 
component (see Table 2), that agree with the predictions of superwind and SBG 
chemical evolution models (Suchkov et al. 1994; Bradamante et al. 1998), favour 
too the galactic wind picture. In such models the galactic wind is powered 
mostly by Type II SNe and minimally by Type Ia SNe: since Type II SNe eject 
mostly $\alpha$-elements and Type Ia SNe eject mostly Fe, the resulting wind is 
overabundant in $\alpha$-elements and underabundant in Fe. It should be noted, 
however, that uncertainties in the absorbing HI column density and the likely 
spread of plasma temperatures may substantially affect the derived chemical 
abundances (see Read et al. 1997; Dahlem et al. 1998). 

The origin of the extended, thermal hard component is less clear.  The
substantial photoelectric absorption obtained from our best-fit requires a
high in-situ hydrogen column density, $N_H \sim 1.3 \times 10^{22}$
cm$^{-2}$ (also the nuclear radio emission suggests the presence of
molecular gas with similar column density: Nakai et al. 1987; Paglione et
al. 1996). This suggests that the hard component may originate in buried,
actively star-forming regions, with Type II SNe and X-ray binaries
possibly supplying the bulk of the X-ray emission and the ambient gas and
dust providing the intrinsic absorption.  In any case, this thermal plasma
cannot be in equilibrium as it is not confined by galactic gravity. To
estimate the contribution of SNe to the observed thermal flux, assuming
that Type II SNe dominate the heating of the ISM, we repeat the swept-up
mass calculation of Ptak et al. (1997) and similarly obtain that $\mincir
50\%$ of the hard component originates in superwind emission and not in
the ambient ISM. The chemical abundances of the hard component, consistent
with those deduced for the soft component, are likewise in agreement with
the trend expected for SB-driven superwind models 
and hence are compatible with the SN origin for (a substantial fraction
of) the hard component. Moreover, Fe could be further depleted
(relative to S and Si) by NGC 253's plentiful dust and warm ISM clouds
(see Telesco 1988). 

However, the galactic wind interpretation of NGC 253's hard component raises 
some basic questions: {\it (i)} what is the mechanism leading to the high 
$\sim 6$ keV temperature?, and {\it (ii)} how many SNRs are required to power 
the hard emission?  In fact: {\it (1)} SNRs usually have substantially lower 
temperatures ($kT \lsimeq 4$ keV); {\it (2)} even neglecting the spectral 
discrepancy, if $L^{\rm SNR}_X \sim 2 \times 10^{35}$ erg s$^{-1}$ then 
$\magcir 8 \times 10^{4}$ SNRs are required to power the hard component of 
NGC 253, while 
$n_{SNR} < 7.5 \times 10^{4} \bigl({R_{SB} \over 5 {\rm kpc}}\bigr)^{2}  
\bigl({z_{SB} \over 0.5 {\rm kpc}}\bigr) \big/ {4 \over 3} 
\bigl({R_{SNR} \over 0.05 {\rm kpc}}\bigr)^{3}$ 
(where $R_{SB}$ and $z_{SB}$ are the radius and thickness of the SB region, 
and $R_{SNR}$ is the SN shock front) SNRs can be at work if the two-phase 
status of NGC 253's ISM is to be preserved. Alternatively, the estimated 
$\sim 4 \times 10^{4}$ SNRs powering the soft component contribute only 
$< 50$\% of the hard component; and {\it (3)} hydrodynamical simulations of 
superwinds do predict hard X-ray emission at such high temperatures, but 
with lower luminosities (compared to the soft X-ray component) and higher 
metal abundances (Suchkov et al. 1994): however, a self-consistent model 
of galaxy formation and evolution, including baryon infall and dissipation, 
star formation, and feed-back onto the still infalling gas, is required to 
test the superwind picture (eg., Heckman et al. 1990) in detail. 

Alternative explanations of the hard component involve either {\it (1)} a 
different mechanism leading to the observed thermal emission or {\it (2)} 
a ``spectral conspiracy'' (discussed in a separate paper) between a softer 
thermal emission and
non-thermal emission yielding the observed spectrum.  As for the former
case, a magnetic confinement picture (e.g., Makishima 1994, 1995) entails
that some fraction of the cooler plasma, confined by magnetic loops in the
galactic disk, is heated to the observed temperature by the dissipation of
the galaxy's rotational and velocity-dispersion energy through magnetic
compression and reconnection.  As for non-thermal spectral components, the
two most likely candidates are X-ray binaries and Compton scattering of
nuclear IR flux by relativistic electrons (Rephaeli et al. 1995; other
nonthermal processes are discussed by Valinia \& Marshall 1998). As concerns 
point-like sources, Ptak et al. (1997), fitting {\it ASCA} SIS spectra for 
two strong extranuclear point sources by a Raymond-Smith plus power-law model, 
argue that the sources' fit parameters are similar to those for the entire 
galaxy, and the two points make up $\sim 25\%$ of the $2-10$ keV flux. While 
some of the point sources in NGC 253 are associated with SNe, some others are 
likely to be X-ray binaries with masses of at least $3-20\,M_\odot$ (implied 
by their Eddington luminosities) suggesting that they are possible blackhole 
candidates. 

Summarizing, the main observational result reported in this Letter is the 
unambiguous detection in NGC 253 of a hot luminous component whose nature is, 
however, still to be clarified.

\acknowledgements{We thank R. Della Ceca and S. Molendi for helpful 
discussion. We also thank the SAX Team for contributing to the operations of 
the satellite and for continuous maintenance of the software.
This research has made use of SAXDAS linearized and cleaned event
files produced at the BeppoSAX Science Data Center. The referee, Andy Ptak,
has helped, with his comments and remarks, to improve the paper.
We acknowledge partial financial support from the Italian Space Agency (ASI) 
and MURST.}

\vglue 0.5truecm

\smallskip
\def\ref{\par\noindent\hangindent 20pt}

\noindent
{\bf 3. References}
\vglue 0.2truecm

\ref{Bhattacharya, D., et al., 1994, ApJ, 437, 173}
\ref{Bookbinder, J., et al. 1980, ApJ 237, 647}
\ref{Bradamante, F., Matteucci, F., \& D'Ercole, A. 1998, A\&A, in press}
\ref{Cappi, M., et al. 1998, to appear in {\it Adv. Space Res.} 
	proceedings of 32$^{\rm nd}$ Scientific Assembly of Cospar}
\ref{Dahlem, M., Heckman, T., \& Fabbiano, G. 1995, ApJ, 442, L49}
\ref{Dahlem, M., Weaver, K.A., \& Heckman, T.M. 1998, ApJS, in press}
\ref{Dickey, J.M. \& Lockman, F.J., 1990, ARA\&A, 28, 215}
\ref{Fabbiano, G., 1988, ApJ, 330, 672}
\ref{Heckman, T.M., Armus, L., \& Miley, G.K. 1990, ApJS, 74, 833}
\ref{Hogg, D.W., et al. 1998 press (astro-ph/9804129)}
\ref{Kaneda, H., et al. 1997, ApJ, 491, 638}
\ref{Madau, P., et al. 1996, MNRAS, 283, 1388}
\ref{Makishima, K. 1994, in New Horizon of X-Ray Astronomy, ed. F.Makino 
	\& T.Ohashi (Tokyo: Universal Academy Press), 171}
\ref{Makishima, K. 1995, in Elementary Processes in Dense Plasmas, ed.
	S.Ichimaru \& S.Ogata (Reading: Addison-Wesley), 47}
\ref{Nakai, N., et al. 1987, PASJ, 39, 685}
\ref{Ohashi, T., et al. 1990, ApJ, 365, 180}
\ref{Paglione, T., Tosaki, T., \& Jackson, J. 1996, ApJ, 454, L117}
\ref{Ptak, A., et al. 1997, AJ, 113, 1286}
\ref{Read, A.M., Ponman, T.J., \& Strickland, D.K. 1997, MNRAS, 286, 626}
\ref{Rephaeli, Y., Gruber, D., \& Persic, M., 1995, A\&A, 300, 91}
\ref{Rieke, G.H. et al. 1980, ApJ, 238, 24}
\ref{Serlemitsos, P., Ptak, A., \& Yaqoob, T. 1996, in ``The Physics of
	Liners'', ed. M.Cracleous et al., 70}
\ref{Suchkov, A.A., et al. 1994, ApJ, 430, 511}
\ref{Telesco, C. 1988, ARA\&A, 26, 343}
\ref{Valinia, A., \& Marshall, F.E. 1998, ApJ, in press (astro-ph/9804012)}
\ref{Yaqoob, T., et al. 1995, ApJ, 455, 508}

\onecolumn
\begin{table}
\begin{center}
\begin{tabular}{ccccccc}
&&&& \\
\multicolumn{7}{c}{Table 3: Two component-thermal models.} \\
\hline
Model & kT$_{soft}$ & Ab$_{soft}$ &  $N_{\rm H}^{\rm hard}$ & kT$_{hard}$ & 
Ab$_{hard}$ & $\chi^2/dof$ \\
 & keV & $\alpha-$el. &  $\times 10^{22}$cm$^{-2}$ & keV & $\alpha-$el. & \\
 & &  Fe, Ni & & & Fe, Ni & \\ 
\hline 
&&&&&& \\
RS & $0.88_{-0.05}^{+0.22}$ & $1.52_{-0.72}^{+0.20}$ & 
$1.28_{-0.20}^{+0.26}$ & $5.62_{-0.20}^{+0.41}$ & $\leq 0.80$ & 246.6/237 \\ 
&&&&&& \\
& & $0.19_{-0.05}^{+0.03}$ & & & $0.25_{-0.06}^{+0.05}$ & \\ 
&&&&&& \\
Mekal & $0.88_{-0.13}^{+0.13}$ & $1.69_{-0.72}^{+0.02}$ & $1.30_{-0.45}^{+0.63}$ & $5.60_{-0.60}^{+0.70}$ & $\leq 1.76$ & 243.6/237 \\
&&&&&& \\
& & $0.18_{-0.07}^{+0.29}$ & & & $0.27_{-0.08}^{+0.10}$ & \\
&&&&&& \\
\hline
\end{tabular}
\end{center}
Notes: The value of the relative normalizations $A_{LECS}/A_{MECS}$ is $\simeq$ 
$0.66$ $\pm$ 0.03. Intervals are at 90\% confidence for one interesting parameter.
\end{table}

\end{document}